\documentclass[aps,prb,twocolumn,showpacs,superscriptaddress,10pt,longbibliography]{revtex4-1}

\usepackage{amsmath}
\usepackage{amssymb}
\usepackage{physics}
\usepackage{graphicx}
\usepackage{subfig}
\usepackage[colorlinks=true,linkcolor=red]{hyperref}
\usepackage[sort&compress]{natbib}
\usepackage{color}

\newcommand{\pcsadd}{Center for Theoretical Physics of Complex Systems, Institute for Basic Science (IBS), Daejeon 34126, Republic of Korea}

\newcommand{\mH}{\mathcal{H}}
\newcommand{\mI}{\mathbb{I}}

\newcommand{\HU}{\mathcal{H}_U}
\newcommand{\dH}{\delta H_1}

\newcommand{\vpsi}{\vec{\psi}}
\newcommand{\bpsi}[1]{\bra{\psi_{#1}}}
\newcommand{\kpsi}[1]{\ket{\psi_{#1}}}
\newcommand{\evpsi}[3]{\mel{\psi_{#1}}{#2}{\psi_{#3}}}

\newcommand{\vphi}{\vec{\varphi}}
\newcommand{\bphi}[1]{\bra{\varphi_{#1}}}
\newcommand{\kphi}[1]{\ket{\varphi_{#1}}}
\newcommand{\evphi}[3]{\mel{\varphi_{#1}}{#2}{\varphi_{#3}}}

\newcommand{\kev}[1]{\ket{e_{#1}}}

\newcommand{\PCLS}{\Psi_\text{CLS}}
\newcommand{\EFB}{E_\text{FB}}

\newcommand{\tw}{\tilde{w}}

\begin{document}

\title{Universal $d=1$ flat band generator from compact localized states}

\author{Wulayimu Maimaiti}
\affiliation{\pcsadd}
\affiliation{Basic Science Program, Korea University of Science and Technology (UST), Daejeon 34113, Republic of Korea}

\author{Sergej Flach}
\affiliation{\pcsadd}

\author{Alexei Andreanov}
\affiliation{\pcsadd}

\date{\today}

\begin{abstract}
    The band structure of some translationally invariant lattice Hamiltonians contains strictly dispersionless flat bands(FB). These are induced by destructive interference, and typically host compact localized eigenstates (CLS) which occupy a finite number $U$ of unit cells. FBs are  important due to macroscopic degeneracy and consequently due to their high sensitivity and strong response to different types of weak perturbations. We use a recently introduced classification of FB networks based on CLS properties, and extend the FB Hamiltonian generator introduced in Phys. Rev. B \textbf{95}, 115135 (2017) to an arbitrary number $\nu$ of bands in the band structure, and arbitrary size $U$ of a CLS. The FB Hamiltonian is a solution to equations that we identify with an inverse eigenvalue problem. These can be solved only numerically in general. By imposing additional constraints, e.g. a chiral symmetry, we are able to find analytical solutions to the inverse eigenvalue problem.
\end{abstract}


\maketitle

\section{Introduction}

Physical models featuring macroscopically degenerate eigenstates have attracted a lot of attention in the past decades. Such degeneracies are naturally unstable to slightest perturbations making them perfect candidates for exotic or unconventional correlated phases of matter like in frustrated magnetism, and strongly correlated systems. An active field in this direction is the understanding of properties of flat bands (FB), i.e., bands with no dispersion~\cite{derzhko2015strongly,leykam2018artificial}. FB models are usually translationally invariant tight-binding networks which are characterized by a certain hopping connectivity between different network sites and which characterize the wave function of, e.g., a quantum particle, a macroscopic condensate, or a photonic field in a structured medium~\cite{leykam2018artificial,leykam2018perspective}. The band structure of the corresponding eigenvalue problem contains $\nu$ bands if the unit cell of the network is containing $\nu$ sites. FB networks were widely studied theoretically in lattice dimension $d=1$~\cite{derzhko2006universal,derzhko2010low,hyrkas2013many}, $d=2$~\cite{mielke1991ferromagnetism,tasaki1992ferromagnetism,misumi2017new}, and in $d=3$~\cite{mielke1991ferromagnetism,nishino2005three,lieb1989two,mielke1991ferromagnetic,mielke1992exact,brandt1992hubbard,ramachandran2017chiral}. FBs have been experimentally realized in a variety of setups, including optical wave guide networks, exciton-polariton condensates, and ultra-cold atomic condensates~\cite{guzman2014experimental,vicencio2015observation,mukherjee2015observation1,weimann2016transport,
xia2016demonstration,taie2015coherent,jo2012ultracold,masumoto2012exciton,baboux2016bosonic}.  

The absence of dispersion in FBs happens due to destructive interference. Destructive interference is also the cause of the existence of \emph{compact localized states} (CLS). CLS are eigenstates at the FB energy, that have strictly finite support on the lattice, and occupy a finite number $U$ of unit cells. Since any translation of a CLS is necessarily again an eigenstate for a translationally invariant Hamiltonian, the existence of a CLS is a direct proof of existence of an FB and its macroscopic degeneracy.

System perturbations typically destroy CLS leading to a variety of interesting phenomena: flat-band ferromagnetism in the fermionic Hubbard model,~\cite{mielke1991ferromagnetic,mielke1991ferromagnetism,tasaki1992ferromagnetism,mielke1993ferromagnetism,tasaki2008hubbard,tasaki1994stability,maksymenko2012flatband} energy dependent scaling of disorder-induced localization length~\cite{leykam2017localization}, singular mobility edges with quasiperiodic potentials~\cite{bodyfelt2014flatbands,danieli2015flatband}, Landau-Zener Bloch oscillations in the presence of external fields~\cite{khomeriki2016landau}, discrete breathers in nonlinear flat band lattices,~\cite{danieli2018breather,johansson2015compactification,real2018controlled}, pair formation of hard core bosons~\cite{mielke2018pair}, and geometric origin of superfluidity~\cite{peotta2015superfluidity,julku2016geometric}. Several approaches were developed to construct FB networks: line graph constructions~\cite{mielke1991ferromagnetism}, decorated lattices~\cite{tasaki1992ferromagnetism}, origami rules~\cite{dias2015origami}, repetition of mini-arrays~\cite{morales2016simple}, chiral symmetry based ones~\cite{ramachandran2017chiral}, and methods based on local symmetries of the Hamiltonian~\cite{roentgen2018compact}. Nishino et al.~\cite{nishino2003flat,nishino2005three} used specific CLS and network symmetries to fine-tune the hoppings down to a FB.

A systematic classification of FBs in terms of compact localized states was introduced in Ref.~\onlinecite{flach2014detangling} where FBs are classified by the size $U$ of the CLS: the number of unit cells occupied by CLS. CLS-based FB generators were then obtained for $U=1$ and arbitrary number of bands and dimension~\cite{flach2014detangling} covering all FB models of that class. For $\nu=2$ and $U=2$ in one dimension, a generator was obtained in Ref.~\onlinecite{maimaiti2017compact} describing all the possible $d=1$ FB networks with two bands. These FB networks form a two-parameter family of generalized sawtooth chains. 

In this work we focus on the case $d=1$ deferring higher dimensions, where we expect even richer phenomenology, for future work. The $d=1$ case was so far analyzed only for two bands and $U=2$ \cite{maimaiti2017compact}. Many recent theoretical proposals \cite{morales2016simple,mondaini2018pairing,gligoric2018nonlinear,murad2018performed,murad2016effective,murad2013geometry, longhi2019photonic,vakulchyk2017anderson} and experimental attempts of realizations \cite{baboux2016bosonic,travkin2017compact,mukherjee2015observation1,weimann2016transport}
focus on $d=1$ settings, and make it necessary to obtain firstly an as complete as possible evaluation of the general $d=1$ case.

We extend the  $\nu=2$ flat band generator \cite{maimaiti2017compact} approach to any value of $\nu$ and $U$. The paper is organized as follow.: In Sec.~\ref{sec:stage} we provide the main definitions that we are using throughout the paper. Sec.~\ref{sec:fb-ieig} discusses the relationship between the FB Hamiltonians and the inverse eigenvalue problems.  That relationship is turned into an efficient FB generator in Sec.~\ref{sec:fb-gen}. In Sec.~\ref{sec:solutions} we present the solutions for the FB generator. We conclude by summarising our results and discussing open problems. 

\section{Main definitions}
\label{sec:stage}

In this work we consider a one-dimensional ($d=1$) translationally invariant lattice Hamiltonian with $\nu>1$ lattice sites per unit cell. We label unit cells by the index $n$, so that the full wave function reads $\Psi=(...,\vpsi_{n-1},\vpsi_n,...)$. Here individual vectors $\vpsi_n$ have elements $\psi_{nm}$, $m=1,...\nu$ labels sites inside the unit cell. Consequently the complex amplitude on the $m$th site in the $n$th unit cell reads as $\psi_{nm}$. We will use the notation $\vpsi_n$ for the wave functions along with the bra-ket notation, $\kpsi{n}$, throughout the paper.

Any translationally invariant Hamiltonian can be characterized by a set of hopping matrices $H_m$, $m=0, 1, \dots$, where $H_0$ is the intracell hopping, $H_1$ describes nearest neighbor unit cell hopping, etc. The case of finite-range hopping is additionally characterized by $m_c$ (the maximum range of the hopping). For the sake of simplicity, we restrict our analysis to the simplest case of $m_c=1$. Most of the results presented below carry over to the cases of $m_c>1$ with minimal changes, that we indicate in the text, where appropriate. We  restrict the analysis to the case of a single flat band in the system, and postpone the more general case of multiple flat bands for later studies.

With the above conventions and notations the eigenvalue problem for an arbitrary nearest-neighbor Hamiltonian reads:~\cite{maimaiti2017compact}
\begin{gather}
    \label{eq:eigen-problem}
    H_1^\dagger\vpsi_{l-1} + H_0\vpsi_l + H_1\vpsi_{l+1} = E\vpsi_l \;.\qquad l\in\mathbb{Z}
\end{gather}

The Hamiltonian of the system is a tri-diagonal block matrix
\begin{gather}
    \label{eq:ham-def}
    \mH = \left(\begin{array}{cccccccc}
        \ddots & \ddots & 0 & 0 & \dots & 0 & 0 & \dots\\
        \ddots & H_0 & H_1 & 0 & 0 & \dots & 0 & \dots\\
        0 & H_1^\dagger & H_0 & H_1 & 0 & \dots & 0 & \dots\\
        \dots & 0 & \ddots & \ddots & \ddots & \ddots & \vdots & \dots\\
        \dots & \vdots & \dots & 0 & H_1^\dagger & H_0 & H_1 & 0\\
        \dots & 0 & \dots & 0 & 0 & H_1^\dagger & H_0 & \ddots\\
        \dots & 0 & 0 & \dots & 0 & 0 & \ddots & \ddots
    \end{array}\right) \;.
\end{gather}

\emph{Cmpact localized state}. A CLS is an eigenvector of~\eqref{eq:eigen-problem} with $\vpsi_n\neq0$ only for a strictly finite number $U$ of adjacent unit cells and zero everywhere else~\cite{flach2014detangling}. The value $U$ is referred to as the \emph{class} of CLS. The presence of a CLS in the spectrum of a translationally invariant Hamiltonian implies an FB. Indeed, in the infinite lattice size limit, infinitely many discrete translations of  a CLS will be linearly independent. A CLS with a larger size $V > U$ can be generated from a given class $U$ CLS by linear superpositions. Therefore the class $U$ refers to the irreducible smallest value of $U$ for which a CLS can not be represented as a linear superposition of even smaller CLS  for a given FB network/Hamiltonian. As far as we can tell, for all known translationally invariant flat band Hamiltonians with finite range hoppings, the FB eigenspace does decompose into a CLS set. For the translationally invariant $d=1$ case the set of all CLS forms a complete basis~\cite{maimaiti2017compact}. The eigenenergy of a flat band will be denoted as $\EFB$. 

The CLS is an eigenvector $\PCLS = (\vpsi_1,\vpsi_2,\dots\vpsi_U)$ of the $U\times U$ block matrix 
\begin{gather}
    \HU = \left(\begin{array}{cccccc}
        H_0 & H_1 & 0 & 0 & \dots & 0\\
        H_1^\dagger & H_0 & H_1 & 0 & \dots & 0\\
        0 & \ddots & \ddots & \ddots & \ddots & \vdots\\
        \vdots & ~ & ~ & ~ & ~ & \vdots\\
        0 & \dots & 0 & H_1^\dagger & H_0 & H_1\\
        0 & \dots & 0 & 0 & H_1^\dagger & H_0
    \end{array}\right)
    \label{eq:cls-def-HU}
\end{gather}
with eigenenergy $\EFB$. Additionally the CLS has to satisfy the destructive interference (compactness) conditions
\begin{equation}
    H_1\vpsi_1 = H_1^\dagger\vpsi_U = 0,
    \label{eq:cls-def-H1}
\end{equation}
that ensure that the wave function amplitudes vanish everywhere except for the $U$ unit cells occupied by $\Psi_\text{CLS}$. \footnote{In the presence of longer-range hopping $m_c > 1$, the CLS compactness conditions become more involved~\cite{maimaiti2017compact} } 
Therefore a \emph{necessary} condition for the existence of a CLS reads
\begin{gather}
    \det H_1 = 0.
\end{gather}

\emph{Chiral symmetry:} An important subclass of FB networks is that with chiral symmetry.~\cite{ramachandran2017chiral}  Chiral lattices are bipartite networks with minority and a majority sublattices. This imposes a specific structure of the hopping integrals and the CLS amplitudes $\vpsi_l$. For that we split the lattice sites from each unit cell into two subsets, each belonging to one of the two sublattices. This leads to a splitting of each $\vpsi_l$ into two sublattice vectors, as well as to a corresponding block structure of the matrices $H_0,H_1$. As a result the CLS of a chiral flat band will always reside exclusively on the majority sublattice~\cite{ramachandran2017chiral}:
\begin{gather}
    H_0 = \left(\begin{array}{cc}
        0 & A^\dagger\\
        A & 0
    \end{array}\right),\quad
    H_1 = \left(\begin{array}{cc}
        0 & T^\dagger\\
        S & 0
    \end{array}\right),\notag\\
    \label{eq:H01-psi-cs-def-b}
    \vpsi_l = \left(\begin{array}{c}
        \vphi_l\\
        0
    \end{array}\right),\quad l=1,\dots,U \;.
\end{gather}
Here, $A$, $S$, and $T$ are $(\nu-\mu)\times \mu$ matrices, $\mu$ is the number of sites on the majority sublattice in the unit cell, and $\vphi_l$ is a $\mu$ component vector residing on the majority sublattice sites in a unit cell. By definition $\nu - \mu\leq\mu < \nu$. The spectrum of the system enjoys particle-hole symmetry around $E=0$. A chiral flat band has energy $\EFB = 0$ and is symmetry protected. For $\nu < 2\mu$ there are $\mu-\lfloor\nu/2\rfloor$ flat bands at $\EFB=0$.~\cite{ramachandran2017chiral} Increasing the range of hopping $m_c > 1$ while preserving the chiral symmetry will keep the chiral flat bands in place. Moreover one can keep the chiral flat bands by partially destroying the chiral and sublattice symmetry. This is achieved by adding hopping terms on the minority sublattice only, since the chiral FB CLS is occupying majority sublattice sites only:
\begin{gather}
    H_0 = \left(\begin{array}{cc}
        0 & A^\dagger\\
        A & B
    \end{array}\right),\quad
    H_1 = \left(\begin{array}{cc}
        0 & T^\dagger\\
        S & W
    \end{array}\right)\notag\\
    \label{eq:H01-psi-cs-def}
    \vpsi_l = \left(\begin{array}{c}
        \vphi_l\\
        0
    \end{array}\right),\quad l=1,\dots,U \;.
\end{gather}
where $B$ and $W$ are $(\nu-\mu)\times(\nu-\mu)$ matrices. Note that the overall particle-hole symmetry of the system is lost, but the original chiral flat bands are still present at $\EFB = 0$. 

\section{The flat band generator}

The flat band generator introduced below is based on a generalization of the concept developed in Ref.~\onlinecite{maimaiti2017compact} for $\nu=U=2$.

\subsection{Inverse eigenvalue problem}
\label{sec:fb-ieig}

We rewrite the CLS problem~[\eqref{eq:cls-def-HU} and \eqref{eq:cls-def-H1}] as

\begin{eqnarray}
%\begin{aligned}
    H_1\vpsi_2 &=& (\EFB - H_0)\vpsi_1 \label{eig-1}, \\
    H_1^\dagger\vpsi_{l-1} + H_1\vpsi_{l+1} &=& (\EFB - H_0)\vpsi_l , 2 \le l \le U-1,
    \label{eig-2}\\
    \label{eig-3}
    H_1^\dagger\vpsi_{U-1} &=& (\EFB - H_0)\vpsi_U,\\
    H_1\vpsi_1 &=& H_1^\dagger\vpsi_U = 0,
    \label{eig-4}\\
    \vpsi_l &=& 0 \;,\;  l<0,\,l>U \;.
    \label{eig-5}
%\end{aligned}
\end{eqnarray}
This set of equations is the starting point of our flat band generator. Our goal is to generate all possible matrices $H_1$ which allow for the existence of a flat band, given a particular choice of $H_0$. Note that $H_0$ can be diagonal (canonical form), but any non-diagonal  Hermitian choice of $H_0$ is fine as well. 

One way to look for solutions is to parametrize $H_1$ and to compute the flat band energy $\EFB$ and the CLS $\PCLS$ for a given set of $U$ and $\nu$. In order to satisfy~\eqref{eig-4} we choose $H_1$ from the space $\mathcal{Z}$ of $\nu \times \nu$ matrices with one zero eigenvalue. Then the directions of the vectors $\vpsi_1,\vpsi_U$ are fixed by the choice of $H_1$, leaving their two norms as free variables. Together with the remaining unknown CLS components and the flat band energy we arrive at $V=(U-2)\nu+3$ variables. The total number of equations from (\ref{eig-1}-\ref{eig-3}) is $E=U\nu$. Since $\nu \geq 2$ it follows that the set of equations is overdetermined. We need $2\nu-3$ additional constraints which will lead us to the proper codimension$(2\nu-3)$ manifold in the space $\mathcal{Z}$. For $\nu=2$, the codimension(1) manifold was computed explicitly and a closed form of the functional dependence of the CLS and flat band energy on $H_1$ was obtained in Ref.~\onlinecite{maimaiti2017compact}. For larger values of $\nu$ (and $U$) the constraint computation turns hard. Therefore we will simply invert the approach--we will define the CLS (thereby setting $U$) and $\EFB$ and generate the proper $H_1$ matrix manifold. This will turn an overcomplete set of equations into an undercomplete one, which is easier to be analyzed.

Let us assume that $\psi_1$ is not orthogonal to $\psi_U$. Multiplying $\langle \psi_U |$ from the left with equation~\eqref{eig-1}, the flat band energy $\EFB$ follows as \footnote{For $m_c>1$, one has to assume $H_m, m<m_c$ are also input parameters}  
\begin{gather}
    \label{eq:efb-def}
    \EFB = \frac{\evpsi{U}{H_0}{1}}{\langle \psi_1\vert \psi_U \rangle} \;.
\end{gather}
For practical purposes we can choose the CLS normalization condition $\langle \psi_1\vert \psi_U \rangle = 1$.  Note that if $\psi_1$ is orthogonal to $\psi_U$, the CLS class is reduced to a $U-1$ class by an appropriate unitary transformation including a redefinition of the unit cell (see Appendix~\ref{app:u-class-reduction}).

We can then treat the problem of flat band generation~\eqref{eig-1}-\eqref{eig-5} as an inverse eigenvalue problem~\cite{boley1987survey}:  given $\EFB$ and $\PCLS$--as well as part of the Hamiltonian--$H_0$, we  reconstruct the Hamiltonian matrix $\mH$, Eq.~\eqref{eq:ham-def}. The idea of finding hopping matrices for a fixed CLS was first introduced by Nishino, Goda, and Kusakabe~\cite{nishino2003flat,nishino2005three}. Our results, even if limited to $d=1$ in the present work, are much more systematic: compared the work of Nishino, Goda, and Kusakabe, we classify CLS by their size $U$, introduce the constraints on $\PCLS$ ensuring that it is a $U$-class CLS and show how to resolve these constraints.

\subsection{The generator}
\label{sec:fb-gen}

We arrive at the following algorithm to construct a Hamiltonian with a flat band from a given CLS.
\begin{enumerate}
    \item Fix the number of bands $\nu$ and the size of the CLS $U$.
    \item Choose $H_0$, either as a diagonal (canonical form), or as any Hermitian matrix.
    \item Choose a real $\EFB$.
    \item Choose $\vpsi_1$ (or $\vpsi_U$).
    \item Exclude $H_1$ from (\ref{eig-1}-\ref{eig-5}), arrive at a set of two linear and further non-linear constraints, and solve them for the remaining CLS components $\vpsi_l$.
    \item Solve the linear system (\ref{eig-1}-\ref{eig-5})  to find $H_1$. 
\end{enumerate}
The system (\ref{eig-1}-\ref{eig-5})  is linear, and therefore it is easy to solve it, or to show that it has no solution. Typically, if this system has a solution, it will be undercomplete and show up with multiple solutions compatible with the input CLS. It is therefore enough to find a \emph{particular} solution $\bar{H}_1$ to Eqs.~(\ref{eig-1}-\ref{eig-5}).  A generic solution $H_1 = \bar{H}_1 + \dH$, where $\dH$ follows from the homogeneous system of equations
\begin{gather}
    \dH\vpsi_2 = 0\notag\\
    \dH^\dagger\vpsi_{l-1} + \dH\vpsi_{l+1} = 0,\ \ \  2 \le l \le U-1\notag\\
    \label{eq:cls-def-ieig-homogeneous}
    \delta H_{U-1}^\dagger\vpsi_{U-1} = 0\\
    \dH\vpsi_1 = \dH^\dagger\vpsi_U = 0\notag\\
    \vpsi_l=0\quad l<0,\,l>U.\notag
\end{gather}
The perturbation $\dH$ is a deformation of the Hamiltonian $\mH$ which preserves the CLS and the flat band energy, and only affects the dispersive part of the spectrum. 

It is also possible to further constrain the network connectivity by choosing specific elements of $H_0$ and/or $H_1$ to be zero. This is easily accounted for in $H_0$, which is an input parameter. The case of $H_1$ is more involved as discussed in Section~\ref{sec:fb-U-gt-3}. 

\section{Solutions}
\label{sec:solutions}

We proceed to classify flat bands in the order of increasing $U$. The $U=1$ case has already been completed in Ref.~\onlinecite{flach2014detangling}, therefore we start our classification with $U=2$.

\subsection{U=2}
\label{sec:fb-U2}

We fix the number of bands to $\nu$, and choose some $H_0$, $\EFB$, and $\kpsi{1}$. The inverse eigenvalue problem Eq.~(\ref{eig-1}-\ref{eig-5}) now reads
\begin{align}
    H_1\kpsi{2} & = \left(\EFB - H_0\right)\kpsi{1}\notag\\
    \bpsi{1}H_1 & = \bpsi{2}\left(\EFB - H_0\right)\notag\\
    \label{eq:cls-ieig-U2-H1}
    H_1\kpsi{1} & = 0\\
    \bpsi{2}H_1 & = 0.\notag
\end{align}
The eigenfunction $\PCLS=(\vpsi_1, \vpsi_2)$ cannot be chosen arbitrarily - its second part $\kpsi{2}$ has to satisfy the following set of linear and non-linear compatibility constraints:
\begin{align}
    \langle \psi_1\vert\psi_2\rangle & = 1\notag\\
    \label{eq:cls-ieig-U2-constraints}
    \evpsi{1}{H_0}{2} & = \EFB\\
    \expval{\EFB - H_0}{\psi_1} & = \expval{\EFB - H_0}{\psi_2}.\notag
\end{align}
The first constraint is simply a choice of normalization of $\PCLS$. The second constraint follows from Eq.~\eqref{eq:efb-def}  and uses $\EFB$ as input variable. The last identity results from multiplying the first equation in Eqs.~\eqref{eq:cls-ieig-U2-H1} by $\bpsi{2}$ from the left, and multiplying the second equation in Eqs.~\eqref{eq:cls-ieig-U2-H1} by $\kpsi{1}$ from the right. It is not possible to solve the third constraint analytically in general, but we present in Appendix~\ref{app:U2-constraints} a numerical algorithm that allows to resolve these constraints and enumerate all the solutions, if existing. If existing, the solution to $\kpsi{2}$ has $\nu-3$ free parameters. For the special case of two bands $\nu=2$, the flat band energy $\EFB$ can not be chosen arbitrarily and needs to be included into the procedure as a to be defined variable. Note that this particular case can be solved in closed analytical form following a different solution strategy\cite{maimaiti2017compact}.

Once $\PCLS=(\vpsi_1, \vpsi_2)$ is known, we can solve Eq.~\eqref{eq:cls-ieig-U2-H1} for $H_1$. First we note that the last two equations - the destructive interference conditions - can be taken into account with the following ansatz for $H_1$: 
\begin{gather}
    \label{eq:cls-ieig-U2-ansatz}
    H_1 = Q_2\,M\,Q_1,\quad Q_i = \mI - \frac{\kpsi{i}\bpsi{i}}{\bpsi{i}\kpsi{i}}.
\end{gather}
Then Eq.~\eqref{eq:cls-ieig-U2-H1} becomes an inverse eigenvalue problem. The details of the derivation are presented in Appendix~\ref{app:ieig-toy} and the solution is
\begin{align}
    H_1 & = G_1 + \delta H_1,\notag\\
    \label{eq:cls-ieig-U2-sol} 
    G_1 & = \frac{\left(\EFB - H_0\right)\kpsi{1}\bpsi{2}\left(\EFB - H_0\right)}{\expval{\EFB - H_0}{\psi_1}} ,\\
    \delta H_1 & = Q_{12} \,K\, Q_{12},\notag
\end{align}

where $K$ is an arbitrary $\nu\times\nu$ matrix and $Q_{12}$ is a joint transverse projector on $\kpsi{1},\kpsi{2}$: $Q_{12}\kpsi{i}=0,\,i=1,2$. If the denominator $\evpsi{1}{\EFB - H_0}{1}\equiv 0$, the above solution is replaced with a more complicated expression involving two different projectors (see Appendix~\ref{app:ieig-toy} for details).

It is instructive to count the number $F$  of free parameters in the above solution, given a fixed $H_0$, $\EFB$ and  $\kpsi{1}$ for $\nu \geq 3$. It is the sum of two contributions: the number of free parameters in $\delta H_1$  and in the particular solution $G_1$, which are $(\nu-2)^2$ and $(\nu-3)$ respectively. The final result is $F=\nu^2-3\nu+1$. It then follows, that the flat band Hamiltonians form a codimension-$(2\nu-2)$ subspace, since $H_0$ is arbitrary, $\dim(H_1)=\nu^2$, and the total number of free parameters at fixed $H_0$ is $F_t=F+1+\nu=\nu^2-2(\nu+1)$. This is a remarkable result, since it shows that flat band Hamiltonians are only weakly fine-tuned, e.g. for $\nu=3$ we find five free parameters when choosing the nine elements of  $H_1$ for an arbitrary chosen $H_0$. Note that the above counting does not apply to the case $\nu=2$ which was studied in Ref.~\onlinecite{maimaiti2017compact} and amounts to two free parameters when choosing the four elements of $H_1$.

\begin{figure}
   \subfloat[]{
        \label{fig:u2_nu3_can}
        \includegraphics[scale=0.35]{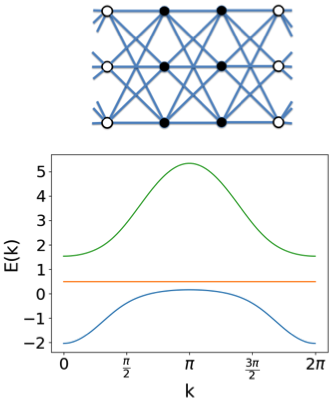}}
   \subfloat[]{
        \includegraphics[scale=0.35]{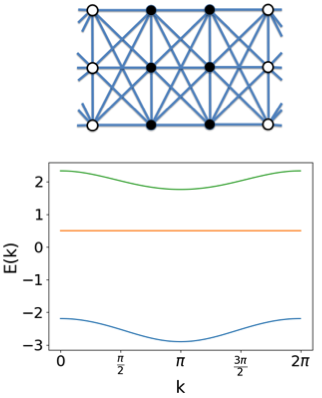}
        \label{fig:u2_nu3_noncan}}\\
   \subfloat[]{
        \includegraphics[scale=0.35]{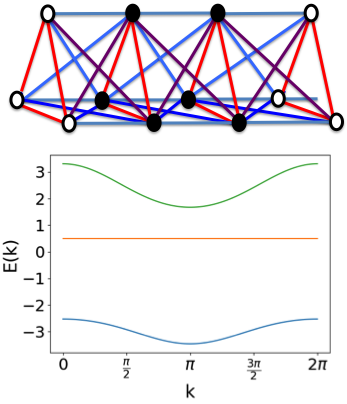}
        \label{fig:u2_nu3_gen}}
    \caption{(Color online) Examples of flat band Hamiltonians with CLS of class $U=2$, $\nu=3$. The sites occupied by a CLS are indicated by filled black circles. Each subfigure contains the visualization of the lattice (top) and the band structure (bottom). The flat band is colored in orange. (a): diagonal $H_0$, (b): non-diagonal $H_0$, (c): non-diagonal and fully connected $H_0$. Appendix~\ref{app:u2-examples} contains the detailed description of the Hamiltonians.}
    \label{fig:u2_examples}
\end{figure}

Equations~\eqref{eq:cls-ieig-U2-constraints} and \eqref{eq:cls-ieig-U2-sol} provide the complete solution to the problem of finding all the $d=1$ nearest-neighbor Hamiltonians with one flat band and CLS of class $U=2$. Figure~\ref{fig:u2_examples} shows some examples of $U=2$ and $\nu=3$ Hamiltonians constructed using the above scheme. 
 
%    Figure 1

\label{sec:fb-U2-cs}

For a bipartite network, the hopping matrix $H_1$ has a specific structure given by Eqs.~\eqref{eq:H01-psi-cs-def}, that simplifies Eqs.~\eqref{eq:cls-ieig-U2-H1} to
\begin{align}
    \label{eq:bipartite_U2_eq}
    S\kphi{2}  & = -A\kphi{1}\\
    S\kphi{1} & = 0\\
    T\kphi{1} & = -A\kphi{2}\\
    T\kphi{2} & = 0,
\end{align}
and $\EFB = 0$. The minority sublattice hopping matrices $B, W$ dropped out as expected. The above equations are considerably simpler than the generic $U=2$ Eqs.~\eqref{eq:cls-ieig-U2-H1}: the above system splits into two independent inverse eigenvalue problems for $S$ and $T$. The details of the solution are presented in Appendix~\ref{app:chiral-inv-eig-prob}, the final answer is
\begin{gather}
    S = -\frac{A\kphi{1}\bphi{2} Q_1}{\mel{\varphi_2}{Q_1}{\varphi_2}} + K_S Q_{12}\notag\\
    \label{eq:cls-ieig-U2-cs-sol}
    T = -\frac{A\kphi{2}\bphi{1} Q_2}{\mel{\varphi_1}{Q_2}{\varphi_1}} + K_T Q_{12},
\end{gather}
where $K_T$ and $K_S$ are arbitrary matrices of size $(\nu-\mu)\times\mu$ respectively. The $Q_{12}$ is a joint transverse projector on $\kphi{1,2}$. There are no restrictions on the entries of $A, B, W$ and $\kphi{1,2}$--they are all free parameters--at variance with the generic $U=2$ flat band construction. Therefore the number of free parameters is: $(\nu-\mu)(2\nu+\mu-2) - 1$ (see Appendix~\ref{app:chiral-inv-eig-prob} for details). The above solution fails for $\mel{\varphi_2}{Q_1}{\varphi_2} = \mel{\varphi_1}{Q_2}{\varphi_1}\equiv 0$, therefore $\kphi{2}\propto\kphi{1}$, the CLS and the flat band are of class $U=1$. 

\begin{figure}
    \includegraphics[scale=0.35]{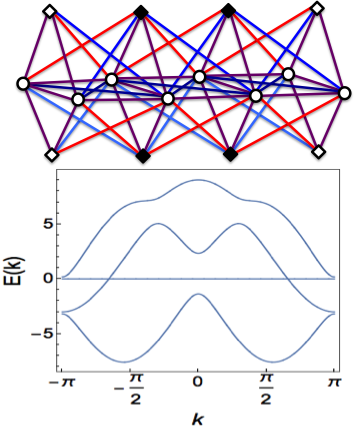}
    \caption{(Color online) Example of a bipartite flat band Hamiltonian with $U=2$, $\nu=4$. The sites of the CLS are indicated by the filled black squares. Links are colored differently for the convenience of visualisation of the chain. In this example the chiral symmetry is broken on the minority sublattice, due to the presence of $B\ne0,W\ne0$ in Eq.~\eqref{eq:H01-psi-cs-def}. Nevertheless the chiral flat band is preserved. The details of this  example are given in Appendix~\ref{app:u2-examples}.}
    \label{fig:u2_bipartite}
\end{figure}

Figure~\ref{fig:u2_bipartite} shows an example of a bipartite lattice with $\nu=4$. There are two sites in the unit cell of each sublattice, and $B\ne0, W\ne0$. In this example, the parameters $\vphi_2,\vphi_2, A, B, W$ are arbitrarily chosen, and $K_T=0, K_S=0$ (See details in the Appendix~\ref{app:u2-examples}).

%    Figure 2

\subsection{$U\ge3$}
\label{sec:fb-U-gt-3}

Let us consider larger $U$ values. For simplicity we use $U=3$ in the examples. Fix the number of bands to $\nu$, and choose some $H_0$, $\EFB$, and $\kpsi{1}$. Then we have the following inverse eigenvalue problem with $U+2$ equations ($U$ for each CLS occupied unit cell, and two for the 
destructive interference conditions):
\begin{align}
    H_1 \kpsi{2} & = \left(\EFB - H_0\right) \kpsi{1}\notag\\
    H_1^\dagger \kpsi{1} + H_1 \kpsi{3} & =\left(\EFB - H_0\right) \kpsi{2}\notag\\
    H_1^\dagger \kpsi{2} & = \left(\EFB - H_0\right) \kpsi{3}\notag\\
    \label{eq:cls-ieig-U3-H1}
    H_1 \kpsi{1} & = 0\\
    H_1^\dagger \kpsi{3} & = 0.\notag
\end{align}

The set of constraints for the $\PCLS$ reads
\begin{gather}
    \langle \psi_{1}\kpsi{3} = 1\notag\\
    \evpsi{1}{H_0}{3} = \EFB\notag\\
    \label{eq:cls-ieig-U3-constraints}
    \evpsi{1}{\EFB - H_0}{2} = \evpsi{2}{\EFB - H_0}{3}\\
    \evpsi{1}{\EFB - H_0}{1} + \evpsi{3}{\EFB - H_0}{3} =\notag\\
     = \evpsi{2}{\EFB - H_0}{2}.\notag
\end{gather}
Again these identities are derived from Eqs.~\eqref{eq:cls-ieig-U3-H1} by multiplying them with $\bpsi{1}$ and  $\bpsi{U}$ and rearranging terms, in order to eliminate $H_1$. Notice that the set of compatibility constraints for $\PCLS$ amounts to $U+1$ equations. Note also that in precisely two of those $U+1$ equations, with $\bpsi{1}$ given, amount to 2 linear, and $U-1$ nonlinear equations for the remaining
CLS amplitudes. It is not possible to solve the nonlinear equations analytically in general, but we present in Appendix~\ref{app:U3-constraints} a numerical algorithm that allows to resolve these constraints and enumerate all the solutions, if existing, for the case $U=3$.

\begin{figure}
    \subfloat[]{\includegraphics[scale=0.37]{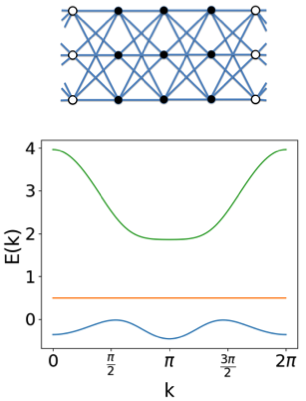}
                        \label{fig:u3_nu3_can}}
    \subfloat[]{\includegraphics[scale=0.37]{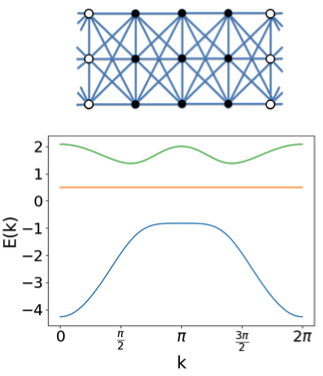}
                        \label{fig:u3_nu3_noncan}}\\
    \subfloat[]{\includegraphics[scale=0.38]{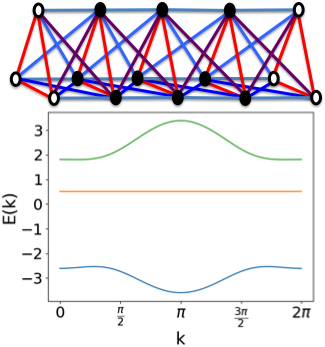}
                        \label{fig:u3_nu3_gen}}
    \subfloat[]{\includegraphics[scale=0.37]{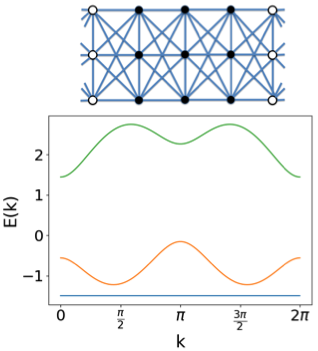}
                        \label{fig:u3_nu3_ground}}
    \caption{ (Color online) Examples of Hamiltonians with a CLS of class $U=3$. The sites occupied by a CLS are indicated by black filled circles. (a) Diagonal choice for $H_0$. (b) Chain like structure for $H_0$. (c) Generic choice for $H_0$. (d) $E_{FB}$ is chosen negative enough to become the groundstate. Details of these examples are presented in Appendix~\ref{app_sec:u3_examples}.}
    \label{fig:u3_examples}
\end{figure}

Instead of using the ansatz~\eqref{eq:cls-ieig-U2-ansatz} for $H_1$, we take a more suitable approach to generate flat band Hamiltonians (i.e. matrices $H_1$) for $U \geq 3$. With a given $\PCLS$ which satisfies the constraints~\eqref{eq:cls-ieig-U3-constraints}, the set of equations~\eqref{eq:cls-ieig-U3-H1} is a linear system with respect to $H_1$:
\begin{gather}
    \label{eq:cls-ieig-U3-H1-linear}
    T\, h_1 = \Lambda.
\end{gather}
Here $h_1$ is a $\nu^2$-dimensional vector resulting from the vectorization of the matrix $H_1$. $T$ is a rectangular $\nu(U+2)\times\nu^2$ matrix whose elements are composed by the elements of CLS, such that the product $T \, h_1$ is the left-hand side of Eqs.~\eqref{eq:cls-ieig-U3-H1}. $\Lambda$  is a $\nu(U + 2)$ vector originating from the right-hand side of Eqs.~\eqref{eq:cls-ieig-U3-H1}:
\begin{gather}
    \label{eq:cls-ieig-U3-lambda}
    \Lambda = (\EFB - H_0)
    \begin{pmatrix}
        \vpsi_1\\
        \vpsi_2\\
        \dots\\
        \vpsi_U\\
        \vec{0}\\
        \vec{0}
    \end{pmatrix}.
\end{gather}
The zero vector components $\vec{0}$ result from the destructive interference. The linear system~\eqref{eq:cls-ieig-U3-H1-linear} can be then solved, e.g., using a least squares solver. Figure~\ref{fig:u3_examples} shows some examples of $U=3$ flat bands, which we generated by resolving the constraints~\eqref{eq:cls-ieig-U3-constraints} and solving Eq.~\eqref{eq:cls-ieig-U3-H1-linear}.

%    Figure 3

\subsection{Network constraints}
\label{sec:lattices}

For practical purposes, the flat band fine-tuning of a Hamiltonian network can involve additional network constraints, e.g. the strict vanishing of certain hopping terms between specific sites of the network~\cite{poli2017partial}. This typically happens when arranging network sites in a plane. Let us consider the typical problem of finding a nearest-neighbor flat band Hamiltonian with specific network constraints. These network constraints dictate the locations of zero entries in $H_0$ and $H_1$. They can be incorporated into the matrix $T$ of Eq.~\eqref{eq:cls-ieig-U3-H1-linear} as a mask $M$: $T \to TM$ that enforces zero entries in $H_1$ in the right positions. The solution of the resulting system is then searched for similar to the non-constrained case.

Especially when  $H_0$ and $H_1$ are sparse , e.g., the number of variables in $H_1$ is equal to or greater then the number of equations, it is possible to solve \eqref{eig-1}-\eqref{eig-5} analytically (see Appendix~\ref{app:imposing_lat_str}). Figure~\ref{fig:imposing_lat_str_examples} shows examples of networks with flat bands generated for a $d=1$ Kagome chain and chains with hoppings allowed only inside network plaquettes.

%    Figure 4
\begin{figure}
    \subfloat[]{
        \includegraphics[scale=0.37]{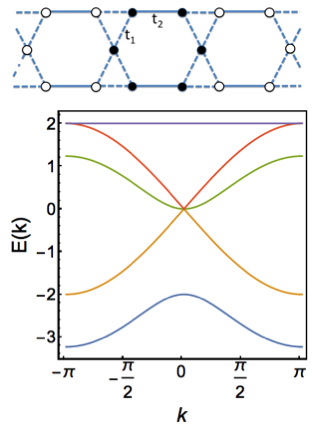}
        \label{fig:1d_kagome}}
    \subfloat[]{
        \includegraphics[scale=0.42]{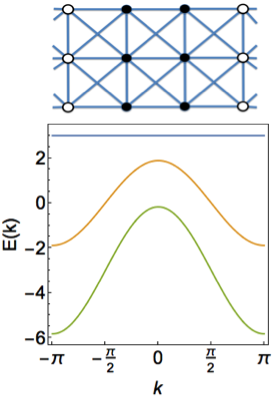}
        \label{fig:u2_nu3_latt_str_example}}\\
    \subfloat[]{
        \includegraphics[scale=0.42]{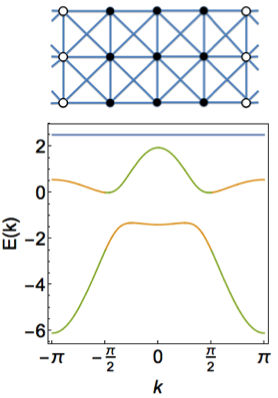}
        \label{fig:u3_nu3_latt_str_example}}
    \caption{(Color online) Examples of flat band  Hamiltonians constructed on specific networks. The sites occupied by a CLS are marked by black filled circles. (a):1d Kagome with $\nu=5$ and $U=2$ compact localized states. The crossing of three bands indicates that the Hamiltonian can be detangled into two independent sub-Hamiltonians. (b) and (c) Examples of Hamiltonians with $\nu=3$, $U=2$ and $U=3$ CLS, respectively. The details of all these Hamiltonians are provided in Appendix~\ref{app:imposing_lat_str}.}
    \label{fig:imposing_lat_str_examples}
\end{figure}

\section{Conclusions}
\label{sec:conclusion}

We presented a systematic construction of one-dimensional Hamiltonians with $\nu$ bands including one flat band for an arbitrary size $U \leq \nu$ of compact localized states and illustrated the method with several examples. The task of finding flat band Hamiltonians is reduced to solving a specific inverse eigenvalue problems subject to certain non-linear constraints. The flat band energy enters as a parameter and can be tuned. For the $U=2$ case we derive analytical solutions to the inverse eigenvalue problem supplemented with a numerical algorithm to resolve for the constraints. For $U \geq 3$ analytical solutions are not accessible, yet numerical algorithms are applied to generate flat band Hamiltonians. We illustrate the method by generating several $U=3$ flat band Hamiltonians. The same construction allows to incorporate various network geometry constraints into the search algorithm. Our results show that flat band Hamiltonians, while being the result of a finetuning in the space of all tight binding Hamiltonian networks, allow for a surprisingly large number of free parameters which change the network, but leave the flatness of the flat band untouched.

Open questions include the extension of the present formalism to the case of multiple flat bands and/or  higher dimensions. The present algorithm can be extended naturally to higher dimensions and will generalize the approach of Nishino \textit{et al}~\cite{nishino2003flat,nishino2005three}. 
The extension to $d=2,3$ would require more intercell hopping matrices $H_a$ describing hopping in different dimensions--$H_x, H_y$ in the simplest case of the square lattice geometry--beyond just $H_1$. Also the simple classification in terms of the CLS size $U$ has to be extended: one has to specify the shape of the compact localised state. Equations~\eqref{eq:eigen-problem} regarded as an inverse eigenvalue problem would now couple different $H_a$. These equations can be decoupled with respect to $H_a$ by introducing additional variables, and reduced to inverse eigenvalue problems for  individual hopping matrices $H_a$, similar to the ones that we were solving here for $d=1$.

Another interesting interesting avenue for future research is the case of non-Hermitian Hamiltonians allowing for gain and loss terms in the Hamiltonian~\eqref{eq:ham-def}. Recently a number of works~\cite{ge2018non,leykam2017flat} analyzed flat bands in such systems or considered the fate of flat bands in the presence of non-Hermitian perturbations~\cite{ge2015parity,longhi2019photonic} and finding interesting results. Finally, non-Hermitian Hamiltonians have a larger parameter space suggesting richer classification as compared to the Hermitian case. We expect therefore that a systematic construction and identification of flat bands in this context might lead to new interesting results.

\begin{acknowledgments}
    This work was supported by the Institute for Basic Science in Korea (IBS-R024-D1).
\end{acknowledgments}

\appendix

\section{Reduction of CLS of class $U$ into $U-1$, when  $\vpsi_1 \perp \vpsi_U$.} 
\label{app:u-class-reduction}

\begin{figure}[h]
    \includegraphics[scale=0.35]{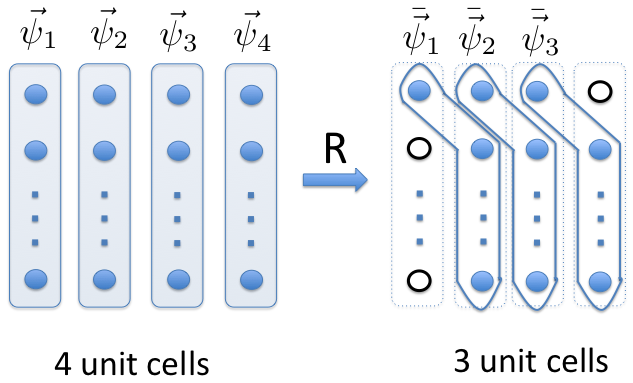}
    \caption{(Color online) A schematics showing how a CLS of class $U=4$ reduces to $U=3$, when $\vpsi_1 \perp \vpsi_4$. Each elongated box stands for one unit cell. Filled circles - nonzero wave function components. Open circles - zero wave function components.}
    \label{fig:app_fig_1}
\end{figure}

Suppose we have a CLS of class $U$, that we write as $\vpsi_{cls} = (\vpsi_1,\vpsi_2,\dots,\vpsi_U)^T$, and $\vpsi_1 \perp \vpsi_U$. Then we can apply a unitary transformation $R$ on the CLS, such that $\tilde{\vpsi}_i=R \vpsi_i,\ i=1,\dots,U$ and
\begin{equation}
  \tilde{\vpsi}_1= \begin{bmatrix}
           1 \\
           0\\
           \vdots \\
           0
         \end{bmatrix}, \quad 
  \tilde{\vpsi}_2= \begin{bmatrix}
           \psi_2^1 \\
           \psi_2^2\\
           \vdots \\
           \psi_2^{\nu}
         \end{bmatrix}, \dots,\quad
  \tilde{\vpsi}_U=\begin{bmatrix}
           0 \\
           \psi_{U}^2\\
           \vdots \\
           \psi_{U}^{\nu}
         \end{bmatrix} , \;
\end{equation}
where $\nu$ is number of sites per unit cell. Due to unitary of transformation $R$, the eigenvalue problem (\ref{eig-1}-\ref{eig-5}) does not change. Next we redefine the unit cell in the following way 
\begin{gather*}
    \bar{\vpsi}_1= \begin{bmatrix}
        1 \\
        \psi_2^2 \\
        \vdots \\
        \psi_2^{\nu}
        \end{bmatrix}, \ 
    \bar{\vpsi}_2= \begin{bmatrix}
        \psi_2^1 \\
        \psi_3^2\\
        \vdots \\
        \psi_3^{\nu}
        \end{bmatrix}, \dots,\ 
    \bar{\vpsi}_{U-1}=\begin{bmatrix}
        \psi_{U-1}^1 \\
        \psi_U^2\\
        \vdots \\
        \psi_U^{\nu}
    \end{bmatrix},
\end{gather*} 
and $\bar{\vpsi}_U = 0$. Therefore, after the unitary transformation $R$ and redefinition of the unit cell, the class of the CLS reduces to $U-1$. The  schematics of this procedure is shown in Figure~\ref{fig:app_fig_1}.

\section{Inverse eigenvalue problem: a toy example and the solution of the $U=2$ CLS}
\label{app:ieig-toy}

This appendix explains the solution of the inverse eigenvalue problems~\eqref{eq:cls-ieig-U2-H1}. As discussed in main text, 1D flat band lattices with CLS class $U$ satisfy 
\begin{align}
    H_1\vpsi_2 & = \left(\EFB - H_0\right)\vpsi_1,\notag\\
    H_1^\dagger\vpsi_{l-1} + H_1\vpsi_{l+1} & = \left(\EFB - H_0\right)\vpsi_l\quad l = 2,\dots,U-1,\notag\\
    H_1^\dagger\vpsi_{U-1} & = \left(\EFB - H_0\right)\vpsi_U,\notag\\
    H_1\vpsi_1 & = 0,\notag\\
    H_1^\dagger\vpsi_U & = 0.
    \label{eq:app-gen-eq}
\end{align}
Assuming that $\EFB, H_0,\vpsi_{l=1,\dots,U}$ are given, the equations~\eqref{eq:app-gen-eq} constitute an inverse eigenvalue problem for a block-tridiagonal matrix, where diagonal blocks are $H_0$ and off diagonal ones are $H_1$. 

\subsection{Toy example}
\label{app:toy}

As a warmup, we solve a toy inverse eigenvalue problem: reconstruct $\nu\times\nu$ matrix $T$ given its action $\ket{y}$ on some vector $\ket{x}$
\begin{gather}
    \label{eq:ieig-toy}
    T\,\ket{x} = \ket{y}.
\end{gather}
The solution is not unique: generic solution can be represented as $T=T_* + \delta T$, where $T_*$ is any particular solution of Eq.~\eqref{eq:ieig-toy} and $\delta T\ket{x}=0$. One possible particular solution is easily found to be
\begin{gather}
    T_* = \frac{\ket{y}\bra{x}}{\bra{x}\ket{x}},\quad\delta T = Q_x\, K,
\end{gather}
where $Q_x$ is a transverse projector on $x$. This construction generalizes straightforwardly to the case of many vectors (we assume here implicitly that the equations are consistent):
\begin{gather}
    T\ket{x_k} = \ket{y_k}\quad k=1..m.
\end{gather}
The generic solution to this problem is given by
\begin{gather}
    T_* = \sum_{ij} A_{ij}\ket{y_i}\bra{x_j},\quad A_{ij}^{-1} = \bra{x_i}\ket{x_j}\\
    \delta T = Q\, K,
\end{gather}
where $Q$ is the orthogonal projector on the subspace spanned by $\{x_k\}$ and $K$ is an arbitrary $\nu\times\nu$ matrix. For later convenience we refer to $T_*$ as \emph{particular solution} and $\delta T$ as \emph{free part}.

\subsection{U=2 case}
\label{app:inv-egv-u2}

In this case, Eq.~\eqref{eq:app-gen-eq} reads
\begin{align}
    H_1\kpsi{2} & = \left(\EFB - H_0\right)\kpsi{1}\notag\\
    H_1^\dagger\kpsi{1} & = \left(\EFB - H_0\right)\kpsi{2}\notag\\
    \label{eq:app-cls-ieig-U2-H1}
    H_1\kpsi{1} & = 0\\
    H_1^{\dagger}\vert\psi_{2}\rangle & = 0.\notag
\end{align}
We know $H_0,\kpsi{1},\kpsi{2}$ and $\EFB=\evpsi{1}{H_0}{2}$, and we need to determine $H_1$. As discussed above for the toy case, the generic solution to this problem can be decomposed into a particular solution and a free part. The last two equations in the above set are satisfied by the following ansatz:
\begin{gather}
    \label{eq:app-cls-ieig-U2-ansatz}
    H_1 = Q_2 M Q_1,\quad Q_i = \mI - \frac{\kpsi{i}\bpsi{i}}{\langle\psi_i\vert\psi_i\rangle}.
\end{gather}
Plugging this ansatz back into the system, we find
\begin{gather}
    \label{eq:app-U2-ieig-M}
    Q_2\, M\,Q_1\kpsi{2} = \left(\EFB - H_0\right)\kpsi{1}\\
    \bpsi{1} Q_2\, M\, Q_1 = \bpsi{2}\left(\EFB - H_0\right).\notag
\end{gather}
Note the identity 
\begin{gather}
    \evpsi{1}{H_1}{2} = \evpsi{1}{\EFB - H_0}{1} = \evpsi{2}{\EFB - H_0}{2},
    \label{eq:app-u2-nl-constraint}
\end{gather}
that follows straightforwardly from the first two equations of~\eqref{eq:app-cls-ieig-U2-H1}. Defining the projectors
\begin{gather}
    R_{12} = \mI - \frac{Q_1\kpsi{2}\bpsi{2} Q_1}{\expval{Q_1}{\psi_2}}\\
    R_{21} = \mI - \frac{Q_2\kpsi{1}\bpsi{1} Q_2}{\expval{Q_2}{\psi_1}},\notag
\end{gather}
we can write 
\begin{equation}
  M = T + R_{21}\, K\, R_{12},
\end{equation}
where $T$ is a particular solution of Eq.~\eqref{eq:app-U2-ieig-M}. The second term, where $K$ is an arbitrary $\nu\times\nu$ matrix, satisfies Eqs.~\eqref{eq:app-U2-ieig-M} by construction and is the free part of the solution. Therefore we only need to find a particular solution to the system to get the generic solution. This is achieved by the same ansatz $T=\ket{x}\bra{y}$ as in the toy case discussed above. The ansatz yields the following equations:
\begin{gather}
    Q_2\, T\, Q_1\kpsi{2} = \mel{y}{Q_1}{\psi_2} Q_2\ket{x} = (\EFB - H_0)\kpsi{1}\\
    \bpsi{1} Q_2\, T\, Q_1 = \mel{\psi_1}{Q_2}{x}\bra{y} Q_1 = \bpsi{2} (\EFB - H_0).
\end{gather}
From these the vectors $x$ and $y$ are fixed (up to unimportant normalization):
\begin{gather*}
    \bra{y} Q_1 = \frac{1}{\mel{\psi_1}{Q_2}{x}}\bpsi{2} (\EFB - H_0)\\
    Q_2\ket{x} = \frac{1}{\mel{y}{Q_1}{\psi_2}} (\EFB - H_0)\kpsi{1}\\
    = \frac{\mel{\psi_1}{Q_2}{x}}{\evpsi{2}{\EFB - H_0}{2}} (\EFB - H_0)\kpsi{1}\\
    = \frac{\mel{\psi_1}{Q_2}{x}}{\evpsi{1}{\EFB - H_0}{1}} (\EFB - H_0) \kpsi{1}.
\end{gather*}
We used the condition~\eqref{eq:app-u2-nl-constraint} to replace the denominator in the fourth line. Also note that the expression for $y$ from the first line was used to simplify the second line, and eliminate $y$. The particular solution is then
\begin{gather}
    \label{eq:app-u2-ansatz-sol}
    Q_2 T Q_1 = \frac{\left(\EFB - H_0\right)\kpsi{1}\bpsi{2}\left(\EFB - H_0\right)}{\evpsi{1}{\EFB - H_0}{1}}
\end{gather}
Thanks to~\eqref{eq:app-u2-nl-constraint} it is symmetric with respect to $\kpsi{1},\kpsi{2}$. This and the above mentioned free part $Q_{21}KQ_{12}$ give the full family of solutions~\eqref{eq:cls-ieig-U2-sol}:
\begin{gather*}
    H_1 = \frac{(\EFB - H_0)\kpsi{1}\bpsi{2}(\EFB - H_0)}{\evpsi{1}{\EFB - H_0}{1}} +  Q_2  R_{21} K R_{12} Q_1
\end{gather*}
This expression is further simplified by noticing that $R_{12}Q_1$ and  $ Q_2 R_{21}$ are the same projector on the subspace spanned by $\kpsi{1},\kpsi{2}$, that we denote $Q_{12}$: $(R_{12}Q_1)^2 = R_{12} Q_1$, idem for $Q_2 R_{21}$ and both vanish when acting on $\kpsi{1,2}$ as can be straightforwardly verified. We can therefore replace these combinations by $Q_{12}$:
\begin{gather}
    \label{eq:app-u2_solution}
    H_1 = \frac{(\EFB - H_0)\kpsi{1}\bpsi{2}(\EFB - H_0)}{\evpsi{1}{\EFB - H_0}{1}} + Q_{12} K Q_{12}.
\end{gather}
This solution is supplemented by the following non-linear constraints
\begin{align}
    \label{eq:app-constraints_u2}
    \langle\psi_2\vert\psi_1\rangle & = 1\\
    \evpsi{2}{H_0}{1}\notag & = \EFB\\
    \evpsi{1}{\EFB - H_0}{1} & = \evpsi{2}{\EFB - H_0}{2},\notag
\end{align}
that are obtained by eliminating $H_1$ from Eq.~\eqref{eq:app-cls-ieig-U2-H1} using "destructive interference conditions", i.e. last two equations in Eq.~\eqref{eq:app-cls-ieig-U2-H1}. 

In case the denominator in Eq.~\eqref{eq:app-u2_solution} is zero, the single projector ansatz fails, and two projector ansatz has to be used:
\begin{equation}
\begin{aligned}
    \label{eq:app-u2_proj2}
    H_1 &= \frac{(\EFB - H_0)\kpsi{1}\bpsi{2}Q_1}{\evpsi{2}{Q_1}{2}} \\
    &+ \frac{Q_2\kpsi{1}\bpsi{2}(\EFB - H_0)}{\evpsi{1}{Q_2}{1}} + Q_{12}K Q_{12},
\end{aligned}
\end{equation}
as can be verified by a direct substitution. In this special solution the denominators only vanish when $\Psi_1\propto\Psi_2$, i.e. in $U=1$ case.

\subsection{Bipartite lattices and chiral symmetry}
\label{app:chiral-inv-eig-prob}

In this section we solve the inverse eigenvalue problem for $U=2$ for the special case of bipartite lattices. We consider a bipartite lattice with $\nu$ sites per unit cell that split into majority and minority sublattices with $\mu$ and $\nu-\mu$ sites respectively. Since the lattice is  bipartite, the sites on one sublattice only have neighbours belonging to the other sublattice. This enforces the following structure on the hopping matrices and the wave functions of the CLS (see Eqs.~\eqref{eq:H01-psi-cs-def}):
\begin{align}
    & H_0 = \left(
    \begin{array}{cc}
        0 & A^\dagger\\
        A & B
    \end{array}\right),\quad
    H_1 = \left(
    \begin{array}{cc}
        0 & T^\dagger\\
        S & W
    \end{array}\right),\notag\\
    & \vpsi_1 = \left(
    \begin{array}{c}
        \varphi_1\\
        0
    \end{array}\right),\quad
    \vpsi_2 = \left(
    \begin{array}{c}
        \varphi_2\\
        0
    \end{array}\right).
    \label{eq:app-chiral-ansatz}
\end{align}
Here $\varphi_{1,2}$ are $\mu$ component vectors describing the wave amplitudes of the majority sublattice sites. $A,S,T$ are $(\nu-\mu)\times\mu$ matrices, while $B, W$ are $(\nu-\mu)\times(\nu-\mu)$ matrices. $B, W$ formally break the bipartiteness of the lattice, but do not affect the $\EFB=0$ flat band(s). This special structure simplifies Eqs.~\eqref{eq:app-cls-ieig-U2-H1}:
\begin{gather}
    S\kphi{2} = -A\kphi{1}\\
    T\kphi{1} = -A\kphi{2}\\
    S\kphi{1} = 0\\
    T\kphi{2} = 0.
\end{gather}
These equations need to be resolved with respect to $S$ and $T$. The last two equations are satisfied by the ans\"atse $S=S^\prime Q_1$, $T=T^\prime Q_2$, where $Q_i$ is a transverse projector on $\varphi_i$. The remaining two equations are identical to the toy problem that we discussed above(see Appendix~\ref{app:toy}) and their solution is precisely Eqs.~\eqref{eq:cls-ieig-U2-cs-sol}:
\begin{gather}
    \label{eq:app-chiral-u2-sol}
    S = -\frac{A\kphi{1}\bphi{2}Q_1}{\evphi{2}{Q_1}{2}} + K_S Q_{12}\\
    T = -\frac{A\kphi{2}\bphi{1}Q_2}{\evphi{1}{Q_2}{1}} + K_T Q_{12},\notag\\
\end{gather}
where $Q_{12}$ is a joint transverse projector on $\kphi{1,2}$.

Now let's count the number of free parameters. $\vert \varphi_1 \rangle , \vert \varphi_2 \rangle$ all are free parameters each contains $\mu$ free parameters. $A$ contains $(\nu-\mu)\mu$ free variables. $B, W$ each contains $(\nu-\mu)^2$ free parameters. $K_S Q_{12}$ and $Q_{21} K_T$ are $(\nu-\mu)\times\mu$ and $\mu\times(\nu-\mu)$ matrices, and, because of the transverse projectors, they contain $(\nu-\mu)(\mu-2)$ and $\mu(\nu-\mu-2)$ free parameters respectively. %Thus $S$ contains $2\mu + (\nu-\mu)(\mu-1)$ free parameters, and $T$ contains $2\mu +\mu(\nu-\mu-1)$.
Therefore total number of free parameters in the solution \eqref{eq:app-chiral-u2-sol} contains $2\mu - 1 + (\nu-\mu)\mu + (\nu-\mu)(\mu-2) + \mu(\nu-\mu-2) + 2(\nu - \mu)^2=(\nu-\mu)(2\nu+\mu-2) - 1$ free parameters.  The extra $-1$ corresponds to the overall normalisation of the CLS, that is not fixed. 

\section{Resolving the non-linear constraints}
\label{app:sol-nl-constraints}

Let us discuss how one can efficiently resolve the set of non-linear constraints, that appear in the inverse eigenvalue problem, for example~\eqref{eq:app-constraints_u2}. Since these are a non-linear system of equations, one can always try a numerical solver. However our experience was not particularly successful: the solver was not converging and finding no solution more often than not. Instead it is possible to design an numerical algorithm that eliminates constraints one by one and either founds and enumerates all the solutions, or proves that there are none.

\subsection{U=2 case}
\label{app:U2-constraints}

The non-linear equations that we need to solve are:
\begin{align}
    \label{eq:app-U2-constraint-l1}
    \langle \psi_1 \vert \psi_2 \rangle & = 1\\
    \label{eq:app-U2-constraint-l2}
    \evpsi{1}{H_0}{2} & = \EFB\\
    \label{eq:app-U2-constraint-l3}
    \expval{\EFB - H_0}{\psi_1} & = \expval{\EFB - H_0}{\psi_2}.
\end{align}
We assume that $\EFB$, $H_0$ and $\psi_1$ (or $\psi_2$) are given input parameters.

Then we need to solve the above equations for $\psi_2$. The first two equations~(\ref{eq:app-U2-constraint-l1}-\ref{eq:app-U2-constraint-l2}) are linear and are easily satisfied with the following expansion for $\psi_2$, by the choice of the basis vectors $e_1$ and $e_2$:
\begin{gather}
    \label{eq:app-U2-psi2-series}
    \kpsi{2} = \sum_{k=1}^\nu x_k \ket{e_k}\\
    \kev{1} = \frac{1}{\sqrt{\bra{\psi_1}\ket{\psi_1}}}\kpsi{1}\\
    \kev{2} = \frac{1}{\sqrt{\evpsi{1}{H_0 Q_1 H_0}{1}}} Q_1 H_0\kpsi{1}\\
    \bra{e_l}\ket{e_m} = \delta_{lm},\quad l,m=1,2,\dots\nu.
\end{gather}
Here $Q_1$ is a transverse projector on $\kpsi{1}$. With this choice of the basis vectors the equations~(\ref{eq:app-U2-constraint-l1}-\ref{eq:app-U2-constraint-l2}) imply:
\begin{gather*}
    x_1 = \frac{1}{\sqrt{\bra{\psi_1}\ket{\psi_1}}},\\
    x_2 = \frac{1}{\sqrt{\evpsi{1}{H_0 Q_1 H_0}{1}}}\left[\EFB - \frac{\evpsi{1}{H_0}{1}}{\bra{\psi_1}\ket{\psi_1}}\right]. 
\end{gather*}

The remaining basis vectors are fixed by requiring their orthonormality, for example, by using Gram-Schmidt orthogonalization. 
Next we plug the expansion~\eqref{eq:app-U2-psi2-series} into Eq.~\eqref{eq:app-U2-constraint-l3} and separate out the terms with $e_1$, $e_2$:
\begin{gather*}
    \expval{\EFB - H_0}{\psi_1} = \sum_{ij=1}^\nu x_i^* x_j\mel{e_i}{\EFB - H_0}{e_j}\\
    = \sum_{ij=1}^2 x_i^* x_j\mel{e_i}{\EFB - H_0}{e_j}\\
    + \sum_{i=1}^2\sum_{j=3}^\nu\left[x_i^* x_j\mel{e_i}{\EFB - H_0}{e_j} + x_j^* x_i\mel{e_j}{\EFB - H_0}{e_i}\right]\\
    + \sum_{ij=3}^\nu x_i^* x_j\mel{e_i}{\EFB - H_0}{e_j}.
\end{gather*}
This expression can be rewritten as follows:
\begin{gather}
    \label{eq:app-U2-nl3-qf}
    \sum_{ij=1}^{\nu-2} y_i^* M_{ij} y_j + \sum_{i=1}^{\nu-2}\left[ u_i^* y_i + u_i y_i^*\right] = w\\
    M_{ij} = \mel{e_{i+2}}{\EFB - H_0}{e_{j+2}}\\
    u_i = \sum_{j=1}^2 x_j \mel{e_{i+2}}{\EFB - H_0}{e_j}\\
    w = \sum_{ij=1}^2 x_i^* x_j\mel{e_i}{\EFB - H_0}{e_j} - \expval{\EFB - H_0}{\psi_1},
\end{gather}
where $y_i = x_{i+2}$. 
The equations on $y_i$ are further simplified by the shift: $z_i=y_i + M_{ij}^{-1} u_j$, that eliminates the linear term. This gives the following equation on a quadratic form
\begin{gather}
    \sum_{ij=1}^{\nu-2} z_i^* M_{ij} z_j = w + \sum_{ij=1}^{\nu-2} u_i^* M_{ij} u_j
\end{gather}
Notice that the RHS of the above equation is real. The matrix $M$ is Hermitian, and can be diagonalized: $M_{ij} = \sum_\alpha E_\alpha \ket{r_\alpha}\bra{r_\alpha}$. The above equation is solved with the help of this spectral decomposition:
\begin{gather}
    \sum_{\alpha=1}^{\nu-2} E_\alpha |t_\alpha|^2 = \tw\\
    \tw = w + \sum_{\alpha=1}^{\nu-2} E_\alpha |s_\alpha|^2\\
    t_\alpha = \bra{r_\alpha}\ket{z_\alpha}\qquad s_\alpha = \bra{r_\alpha}\ket{u}.
\end{gather}
The presence or absence of solution is decided by the mutual signs of $\tw$ and $E_\alpha$: if $\tw>0$ and $E_\alpha<0$ $\forall\alpha$, then there is no solution. If one $E_\alpha > 0$, there is a single solution, for two and more $E_\alpha>0$ there is a multiparametric family of solutions. Knowing $t_\alpha$, it is straightforward to reconstruct the original $\vpsi_2$.

In the above $M$ was assumed non-singular. If it is singular, than $M_{ij}^{-1}$ is the Moore-Pensrose pseudoinverse~\cite{ben2003generalized} and we have $y_i = z_i + g_i - M_{ij}^{-1} u_j$ where $g\in\ker{M}$. For $g_i$ the quadratic terms in~\eqref{eq:app-U2-nl3-qf} vanish (by definition of $g_i$) and the $g_i$ only enter linearly the equation, while $z_i$ can be treated as in the non-singular case (for convenience we assume that the first $k$ eigenvalues of $M$ are zero):
\begin{gather}
    \sum_{\alpha=k+1}^{\nu-2} E_\alpha t_\alpha^2 = \tw - \sum_{\alpha=1}^{k} \left[\bra{u}\ket{r_\alpha} + \bra{r_\alpha}\ket{u}\right].
\end{gather}
The presence of zero modes renormalizes $\tw$.

The more refined version of the counting relies on the above solution, and the counting of the $E_\alpha$ with the ``right" sign. It tells us that for $\nu=2, 3$, there is a single solution for fixed $\vpsi_1,\EFB,H_0$. For larger $\nu$, there could be a single solution or multiparametric families of solutions, from $0$ to $\nu-3$.

\subsection{ U=3 case}
\label{app:U3-constraints}

In this case the nonlinear constraints read, Eq.~\eqref{eq:cls-ieig-U3-constraints}:
\begin{align}
    & \langle\psi_1\vert\psi_3\rangle = 1\notag\\
    & \evpsi{1}{H_0}{3} = \EFB\notag\\
    & \evpsi{1}{\EFB - H_0}{2} = \evpsi{2}{\EFB - H_0}{3}
    \label{eq:app-u3-cls-constraint}\\
    & \evpsi{3}{\EFB - H_0}{3} = \evpsi{2}{\EFB - H_0}{2}\notag\\
    & - \evpsi{1}{\EFB - H_0}{1}\notag
\end{align}
The resolution of this set of constraint is very similar to the $U=2$ case, therefore we only outline the main steps. We search to resolve the above equations with respect to $\Psi_3$, taking $\Psi_1,\Psi_2$ as inputs. The first $3$ equations are linear, and we solve them by expanding $\Psi_3$ over a suitable orthonormal basis:

\begin{align*}
    \kpsi{3} & = \sum_k x_k \kev{k},\\
    \kev{1} & = \frac{1}{\sqrt{\bra{\psi_1}\ket{\psi_1}}} \kpsi{1},\\
    \kev{2} & = \frac{1}{\sqrt{\evpsi{1}{H_0 Q_1 H_0}{1}}} Q_1 H_0\kpsi{1},\\
    \kev{3} & = \frac{Q_*(\EFB - H_0)\kpsi{2}}{\sqrt{\evpsi{2}{(\EFB - H_0) Q_* (\EFB - H_0)}{2}}},\\
    & \vdots\\
    \bra{e_l}\ket{e_m} & = \delta_{l,m},\ \ l,m=1,2,\dots,\nu\\
    Q_1 & = \mathbb{I} - \frac{\kpsi{1}\bpsi{1}}{\bpsi{1}\kpsi{1}},
\end{align*}
and $Q_*$ is a joint transverse projector on $\kpsi{1}$ and $Q_1 H_0\kpsi{1}$. Then $x_1$ and $x_2$ are the same as in the $U=2$ case, $x_3$ is directly expressed through the third equation in Eqs.~\eqref{eq:app-u3-cls-constraint}. The last, fourth equation in~\eqref{eq:app-u3-cls-constraint} is solved in the same way as that in the $U=2$ case: it is reduced to solving a quadratic form.

\section{Examples for FB generators}
\label{app:examples}

In this section we present the details of the example flat band Hamiltonians generated using the algorithm discussed in the main text. In all of these examples we pick some $H_0, \EFB$ and part of the $\psi_l$ as an input. Next following the algorithm outlined in the Appendix~\ref{app:sol-nl-constraints} we construct a  set of $\{\psi_l\}$ consistent with the CLS structure. Then we find the hopping matrix $H_1$ using the algorithm from Section~\ref{sec:fb-gen} (detailed in Appendix~\ref{app:ieig-toy}). For simplicity we drop the free part $K$ in all the examples below.

\subsection{$\nu=3, U=2$ case}
\label{app:u2-examples}

\emph{Example shown in Fig.~\ref{fig:u2_nu3_can}}: We start with a three band case $\nu=3$, and no additional constraints on the form of $H_1$. We assume canonical (diagonal) form of
$H_0$ and choose $\vpsi_1$ 
\begin{gather}
    H_0 = \left[\begin{array}{ccc}
        0 & 0 & 0\\
        0 & 1 & 0\\
        0 & 0 & 2
    \end{array}\right],
    \ \ \vpsi_1 = \left(1, -1, 1\right),
\end{gather}

Using the FB algorithm, we find the particular solution:
\begin{align}
    & \EFB = 0.5,\quad\vpsi_2 = \left(1.5,\ 1.5,\ 1\right)\\
    & H_1 = \left[\begin{array}{ccc}
        -0.25 &  0.25 &  0.5 \\
        -0.25 &  0.25 & 0.5\\
        0.75 & -0.75 & -1.5
    \end{array}\right], 
\end{align}

\emph{Example shown in Figure~\ref{fig:u2_nu3_noncan}}: Taking non-diagonal $H_0$ and $\vpsi_1$ as 
\begin{gather}
    H_0 = \left[\begin{array}{ccc}
        0 & 1 & 0\\
        1 & 0 & 1\\
        0 & 1 & 0
    \end{array}\right],
    \ \ \vpsi_1 = \left( 1, -1,  1\right),
\end{gather}

we construct the following FB Hamiltonian:
\begin{align}
    & H_1 = \left[\begin{array}{ccc}
        0.19926929 & -0.47727273 & -0.67654202\\
        -0.33211549 &  0.79545455 &  1.12757003\\
        0.19926929 & -0.47727273 & -0.67654202
    \end{array}\right],\\ 
    & \vpsi_2 = \left(4.46130814,\ 1.5,\  -1.96130814\right),\ \EFB = 0.5
\end{align}

\emph{Example shown in Figure~\ref{fig:u2_nu3_gen}}: Taking all the sites in the unit cell connected to each other and the same $\vpsi_1$, $\EFB$ as in the above example

\begin{gather}
    H_0 = \left[\begin{array}{ccc}
        0 & 1 & 1\\
        1 & 0 & 1\\
        1 & 1 & 0
    \end{array}\right],\
    \ \vpsi_1 = \left(1,-1,1\right)
\end{gather}
we land at the following Hamiltonian:
\begin{align}
    & H_1 = \left[\begin{array}{ccc}
        0.18163216 & -0.16071429 & -0.34234644\\
        -0.90816078 &  0.80357143 &  1.71173221\\
        0.18163216 & -0.16071429 & -0.34234644
    \end{array}\right],\\
    & \vpsi_2 = \left(1.84761673,\ 0.25 ,\ -0.59761673\right),\ \EFB = 0.5
\end{align}

\emph{Bipartite lattice $U=2$, Figure~\ref{fig:u2_bipartite}}: 
We consider the $\nu=4, \mu=2$ case and pick the following input variables:
\begin{align*}
    & A = \frac{1}{4} \left(\begin{array}{cc}
        \sqrt{3} & 1 \\
        3 & \sqrt{3} \\ \end{array} \right), \ \ 
    B=\left(\begin{array}{cc}
        1 & -2 \\
        -2 & 1 \\ \end{array} \right)\\ 
    & \vphi_1 = \frac{1}{\sqrt{2}} \left( 1, 1 \right),
    \ \ \vphi_2 = \frac{1}{2} \left( 1, \sqrt{3} \right) \\
    & W = \left(\begin{array}{cc}
        2 & -1 \\
        -1 & 2 \\
    \end{array}\right).
\end{align*}
Solving Eqs.~\eqref{eq:cls-ieig-U2-cs-sol}/\eqref{eq:app-chiral-u2-sol} yields the following solution: 
\begin{align*}
    & S = \frac{\left(\sqrt{3}+2\right)
    \left(\begin{array}{cc}
        -1 & 1 \\
        -\sqrt{3} & \sqrt{3} \\
    \end{array}\right)}{2 \sqrt{2}}, \\
    & T = \frac{\left(\sqrt{3}+1\right)
    \left(\begin{array}{cc}
        3 & 3 \sqrt{3} \\
        -\sqrt{3} & -3 \\
    \end{array}\right)}{4 \sqrt{2}}
\end{align*}
The corresponding hopping matrices $H_0, H_1$ read
\begin{align*}
    H_0 & =  \left(\begin{array}{cccc}
        0 & 0 & \frac{\sqrt{3}}{4} & \frac{3}{4} \\
        0 & 0 & \frac{1}{4} & \frac{\sqrt{3}}{4} \\
        \frac{\sqrt{3}}{4} & \frac{1}{4} & 1 & -2 \\
        \frac{3}{4} & \frac{\sqrt{3}}{4} & -2 & 1 \\
    \end{array}\right)\\
    H_1 & = \left(\begin{array}{cccc}
        0 & T  \\
        S & W \\
    \end{array}\right)\\
    \vpsi_1 & = \left(\frac{1}{\sqrt{2}},\frac{1}{\sqrt{2}},0,0\right) \\
    \vpsi_2 & = \left(\frac{1}{2},\frac{\sqrt{3}}{2},0,0\right).
\end{align*}

\subsection{$\nu=3$, $U=3$ case}
\label{app_sec:u3_examples}

\emph{Example shown in Figure~\ref{fig:u3_nu3_can}}: We pick $H_0$ in canonical form and choose $\vpsi_1$ as follows 
\begin{gather*}
    H_0 = \left[\begin{array}{ccc}
        0 & 0 & 0\\
        0 & 1 & 0\\
        0 & 0 & 2
    \end{array}\right],
    \ \ \vpsi_1 = \left(1, -1, 1\right)
\end{gather*}
Solving the non-linear constraints~\eqref{eq:cls-ieig-U3-constraints}/\eqref{eq:app-u3-cls-constraint}, we get $\vpsi_2, \vpsi_3$. Then solving the equation~\eqref{eq:cls-ieig-U3-H1-linear}, which is equivalent to equations (\ref{eig-1}-\ref{eig-5}), we get
\begin{align*}
    H_1 & = \left[\begin{array}{ccc}
        -0.06548573 & -0.27210532 & -0.2066196 \\
        -0.15130619 & -0.28682832 & -0.13552213\\
        -0.14682469 &  0.75742396 &  0.90424865
    \end{array}\right]\\
    \vpsi_2 & = \left(-0.05144152, -1.53640189, -0.38025523\right)\\
    \vpsi_3 & = \left(0.58333333, -0.33333333,  0.08333333\right)\\
    \EFB & = 0.5
\end{align*}

\emph{Example shown in Figure~\ref{fig:u3_nu3_noncan}}: We choose $H_0$ and $\vpsi_1$ as 
\begin{gather*}
    H_0 = \left[\begin{array}{ccc}
        0 & -1 & 0\\
        -1 & 0 & 1\\
        0 & 1 & 0
    \end{array}\right],
    \ \ \vpsi_1 = \left(1, -1, 1\right)
\end{gather*}
The corresponding flat band $H_1$ is
\begin{align*}
    H_1 & = \left[\begin{array}{ccc}
        0.23624218 &  0.15535892 & -0.08088326\\
        -0.87350793 & -0.69073091 &  0.18277702\\
        1.31303601 &  0.95651792 & -0.35651809
    \end{array}\right]\\
    \vpsi_2 & = \left(3.14189192, -2.05220768, -0.94681365\right)\\
    \vpsi_3 & =\left(1.08333333, -0.33333333, -0.41666667\right)\\
    \EFB & = 0.5
\end{align*}

\emph{Example shown in Figure~\ref{fig:u3_nu3_gen}}: For the following input
\begin{gather*}
    H_0 = \left[\begin{array}{ccc}
        0 & -1 & 2\\
        -1 & 0 & 1\\
        2 & 1 & 0
    \end{array}\right],
    \ \ \vpsi_1 = \left(1, -1, 1\right),
\end{gather*}
we find the flat band n.n. hopping matrix $H_1$:
\begin{align*}
    H_1 & = \left[\begin{array}{ccc}
        0.06915801 & -0.66620419 & -0.7353622\\
        -0.31644957 & -0.3029663 &   0.01348327\\
        -0.46657738 & -0.38011423 &  0.08646314
    \end{array}\right]\\
    \vpsi_2 & =\left(0.77717503,  2.50899893,  1.05355773\right)\\
    \vpsi_3 & =\left(0.03571429, -0.57142857,  0.39285714\right)\\
    \EFB & =0.5.
\end{align*}

\emph{Example shown in Figure~\ref{fig:u3_nu3_ground}}: The following input data
\begin{gather*}
    H_0 = \left[\begin{array}{ccc}
        0 & 1 & 0\\
        1 & 0 & 1\\
        0 & 1 & 0
    \end{array}\right],
    \ \ \vpsi_1 = \left(1, -1, 1\right),
\end{gather*}
provides an example of the flat band Hamiltonian, with the flat band being the ground state:
\begin{align*}
    H_1 & = \left[\begin{array}{ccc}
        -0.52279625 &  0.17024672 &  0.69304298\\
        -0.62702148 & -0.11461122 &  0.51241027\\
        -0.73124671 & -0.39946915 &  0.33177756
    \end{array}\right]\\
    \vpsi_2 & = \left(0.25537008,  0.28652804, -0.59920373\right)\\
    \vpsi_3 & = \left(0.25, -0.5,   0.25\right)\\
    \EFB & =-1.5.
\end{align*}

\section{Network constraints}
\label{app:imposing_lat_str}

We present here the details of the examples where the network connectivity was provided as an input to the FB generator. In all cases one can find particular solutions to the resulting non-linear system of equations.

Often network connectivity implies sparse $H_0$ and $H_1$ very sparse. Therefore inserting these sparse $H_0$ and $H_1$ into equations (\ref{eig-1}-\ref{eig-5}) gives a set of equations that can be solved analytically. More precisely, as you will see in the examples below, when $H_0$ and $H_1$ are so sparse that the number unknowns (non-zero elements of $H_1, H_0$ and part of CLS) is less then or equal to the number of equations, we can solve the equations (\ref{eig-1}-\ref{eig-5}) analytically. Note that, instead of inserting $H_1$ and $H_0$ into equations (\ref{eig-1}-\ref{eig-5}), we can get the same set of equations from equation~\eqref{eq:cls-ieig-U3-H1-linear} by zeroing the elements of $h_1$ corresponding to zero elements of $H_1$.

\subsection{U=2 Case}

\subsubsection{1D Kagome}

We consider the $d=1$ version of the 2D Kagome lattice. The n.n. Hamiltonian is restricted by the lattice connectivity to
\begin{gather*}
    H_0 = \left[\begin{array}{ccccc}
        0 & t_2 & 0 & 0 & 0\\
        t_2 & 0 & t_1 & 0 & 0\\
        0 & t_1 & 0 & t_1 & 0\\
        0 & 0 & t_1 & 0 & t_2\\
        0 & 0 & 0 & t_2 & 0
    \end{array}\right],
    \ \ H_1 = \left[\begin{array}{ccccc}
        0 & t_1 & t_1 & 0 & 0\\
        0 & 0 & 0 & 0 & 0\\
        0 & 0 & 0 & 0 & 0\\
        0 & 0 & 0 & 0 & 0\\
        0 & 0 & t_1 & t_1 & 0
    \end{array}\right]
\end{gather*}
The "\emph{destructive interference}" condition~\eqref{eq:cls-def-H1} ,i.e. the last two equations in \eqref{eq:cls-ieig-U2-H1},
implies that 
\begin{gather*}
    \vpsi_1 = \left(x_1, -x_2, x_2, -x_2, x_3\right),
    \ \ \vpsi_2 = \left(0, a, b, c, 0\right).
\end{gather*}
 If we insert $\vpsi_1,\vpsi_2$ above into the equations \eqref{eq:cls-ieig-U2-H1}, we find 
\begin{align*}
\left(\begin{array}{c}
-x_{2}t_{2}+\left(y_{2}+y_{3}\right)t_{1}\\
x_{2}t_{1}+x_{1}t_{2}\\
-2x_{2}t_{1}\\
x_{2}t_{1}+x_{3}t_{2}\\
-x_{2}t_{2}+\left(y_{3}+y_{4}\right)t_{1}
\end{array}\right) & =E_{FB}\left(\begin{array}{c}
x_{1}\\
-x_{2}\\
x_{2}\\
-x_{2}\\
x_{3}
\end{array}\right)\\
\left(\begin{array}{c}
at_{2}\\
\left(b+y_{1}\right)t_{1}\\
\left(a+c+y_{1}+y_{5}\right)\\
\left(b+y_{5}\right)t_{1}\\
ct_{2}
\end{array}\right) & =E_{FB}\left(\begin{array}{c}
0\\
a\\
b\\
c\\
0
\end{array}\right)
\end{align*}
One the possible solutions of above equation is 
\[
\begin{aligned}a=c & =0\\
t_{1} & =-\frac{E_{FB}}{2}\\
t_{2} & =\frac{E_{FB}}{2}\\
x_{1} & =-x\\
x_{2} & =x\\
x_{3} & =-x\\
a & =0\\
b & =x\\
c & =0
\end{aligned}
\]
 This solution gives a flat band with energy $E_{FB}$. Thus the final
solution is 
\[
\begin{aligned}\vec{\psi}_{1} & =\left(-x,-x,x,-x,-x\right)\\
\vec{\psi}_{2} & =\left(0,0,x,0,0\right)\\
H_{1} & =\left[\begin{array}{ccccc}
0 & -\frac{E_{FB}}{2} & -\frac{E_{FB}}{2} & 0 & 0\\
0 & 0 & 0 & 0 & 0\\
0 & 0 & 0 & 0 & 0\\
0 & 0 & 0 & 0 & 0\\
0 & 0 & -\frac{E_{FB}}{2} & -\frac{E_{FB}}{2} & 0
\end{array}\right]\\
H_{0} & =\left[\begin{array}{ccccc}
0 & \frac{E_{FB}}{2} & 0 & 0 & 0\\
\frac{E_{FB}}{2} & 0 & -\frac{E_{FB}}{2} & 0 & 0\\
0 & -\frac{E_{FB}}{2} & 0 & -\frac{E_{FB}}{2} & 0\\
0 & 0 & -\frac{E_{FB}}{2} & 0 & \frac{E_{FB}}{2}\\
0 & 0 & 0 & \frac{E_{FB}}{2} & 0
\end{array}\right]
\end{aligned}
\]
 This lattice has a flat band with flat band energy $E_{FB}$.

\subsubsection{$U=2$, $\nu=3$ example}
\label{sec:app-imposing-lat-str-u2-nu3}

The connectivity of the network shown in Figure~\ref{fig:u2_nu3_latt_str_example} implies the following hopping matrices:
\begin{align*}
    & H_0 = \left(\begin{array}{ccc}
        0 & t_1 & 0 \\
        t_1 & 0 & t_2 \\
        0 & t_2 & 0 \\
    \end{array}\right), \ \
    H_1 = \left(\begin{array}{ccc}
        s_1 & s_2 & 0 \\
        s_4 & s_5 & s_6 \\
        0 & s_7 & s_8 \\
    \end{array}\right)
\end{align*}
We parameterize the CLS amplitudes as follows: $\vpsi_1 = (x,y,z),\vpsi_2 = (a,b,c)$. Then equations \eqref{eq:cls-ieig-U2-H1} gives 
\begin{align*}
    & \left(\begin{array}{c}
        a s_1+b s_2 \\
        a s_4+b s_5+c s_6\\
        b s_7+c s_8
    \end{array}\right)=
    \left(\begin{array}{c}
        x \EFB  - t_1 y\\
        -t_1 x-t_2 z+ y \EFB\\
        z \EFB - t_2 y
    \end{array}\right)\\
    & \left(\begin{array}{c}
        s_1 x+s_4 y \\
        s_2 x+s_5 y+s_7 z \\
        s_6 y+s_8 z
    \end{array}\right) =
    \left(\begin{array}{c}
        a \EFB - b t_1 \\
        -a t_1 + b \EFB - c t_2 \\
        c \EFB - b t_2
    \end{array} \right) \\
    & \left(\begin{array}{c}
        s_1 x+s_2 y \\
        s_4 x+s_5 y+s_6 z \\
        s_7 y+s_8 z
    \end{array}\right) = 
    \left(\begin{array}{c}
        0 \\
        0 \\
        0
    \end{array} \right) \\
    & \left(\begin{array}{c}
        a s_1+b s_4 \\
        a s_2+b s_5+c s_7 \\
        b s_6+c s_8
    \end{array}\right)=
    \left(\begin{array}{c}
        0 \\
        0 \\
        0
    \end{array} \right)
\end{align*}
Here $H_0$, $\EFB$ and $\vpsi_1$ as free parameters. If we fix $x=1,y= 2,z=1,t_1=1,t_2= 2,\EFB = 3$, then we find one particular solution of above equations
\begin{align*}
    & s_1=\frac{2 \sqrt{2}}{3}, \ \ s_2 = -\frac{\sqrt{2}}{3} \\
    & s_4 = \frac{2 \sqrt{2}}{3}, \ \  s_5 = \frac{\sqrt{2}}{3} \\
    & s_6 = -\frac{1}{3} \left(4 \sqrt{2}\right), \ \  s_7 = -\frac{\sqrt{2}}{3}, \\
    & s_8 = \frac{2 \sqrt{2}}{3}, \ \ a = \frac{1}{\sqrt{2}} \\
    & b = -\frac{1}{\sqrt{2}}, \ \ c = -\sqrt{2}
\end{align*}
from which follow the hopping matrices and the CLS amplitudes
\begin{align*}
    & H_0 = \left(\begin{array}{ccc}
        0 & 1 & 0 \\
        1 & 0 & 2 \\
        0 & 2 & 0 \\
    \end{array}\right), \ \ 
    H_1 = \left(\begin{array}{ccc}
        \frac{2 \sqrt{2}}{3} & -\frac{\sqrt{2}}{3} & 0 \\
        \frac{2 \sqrt{2}}{3} & \frac{\sqrt{2}}{3} & -\frac{1}{3} \left(4 \sqrt{2}\right) \\
        0 & -\frac{\sqrt{2}}{3} & \frac{2 \sqrt{2}}{3} \\
    \end{array}\right) \\
    & \vpsi_1 = (1, 2, 1),\ \ \vpsi_2 = \left(\frac{1}{\sqrt{2}},-\frac{1}{\sqrt{2}},-\sqrt{2} \right).
\end{align*}
%The schematics and band structure of this lattice is shown if figure \ref{fig:u2_nu3_latt_str_example}.

\subsection{ U=3 case}

\subsubsection{$U=3$, $\nu=3$ example}
\label{sec:app-imposing-lat-str-u3-nu3}

We consider networks shown in Fig.~\ref{fig:u3_nu3_latt_str_example}. Its connectivity requires the following hopping matrices
\begin{gather*}
    H_0 = \left(\begin{array}{ccc}
        0 & t_1 & 0 \\
        t_1 & 0 & t_2 \\
        0 & t_2 & 0 \\
    \end{array}\right), \ \ 
    H_1 =\left(\begin{array}{ccc}
        s_1 & s_1 & 0 \\
        -\frac{s_1}{2} & -\frac{s_1}{2}-s_6 & s_6 \\
        0 & 2 s_6 & -2 s_6 \\
    \end{array}\right)
\end{gather*}
According to "\emph{destructive interference}" condition~\eqref{eq:cls-def-H1}, we paramterize $\vpsi_1,\vpsi_2,\vpsi_3$ as follows 
\begin{gather*}
    \vpsi_1 = (-y, y, y), \ \ \vpsi_2 = (a,b,c), \ \ \vpsi_3 = (d, 2d, d)
\end{gather*}
Then the main equations~\eqref{eq:cls-ieig-U3-H1} become:
\begin{align*}
    & \left( \begin{array}{c}
        (a+b) s_1 \\
        (c-b) s_6-\frac{1}{2} (a+b) s_1 \\
        2 (b-c) s_6 \\
    \end{array} \right) =
    \left(\begin{array}{c}
        -y \left(\EFB + t_1\right) \\
        y \left(\EFB + t_1-t_2\right) \\
        y \left(\EFB - t_2\right) \\
    \end{array} \right) \\
    & \left(\begin{array}{c}
        \frac{3}{2} (2 d-y) s_1 \\
        (y-d) s_6-\frac{3}{2} (d+y) s_1 \\
        (2 d-y) s_6 \\
    \end{array} \right) =
    \left(\begin{array}{c}
        a \EFB - b t_1 \\
        b \EFB - a t_1-c t_2 \\
        c \EFB - b t_2 \\
    \end{array} \right) \\
    & \left(\begin{array}{c}
        \frac{1}{2} (2 a-b) s_1 \\
        \left(a-\frac{b}{2}\right) s_1-(b-2 c) s_6 \\
        (b-2 c) s_6 \\
    \end{array} \right) =
    \left( \begin{array}{c}
        d \left(\EFB - 2 t_1\right) \\
        d \left(2 \EFB - t_1 - t_2\right) \\
        d \left(\EFB - 2 t_2\right) \\
    \end{array} \right)
\end{align*}
%Considering $t_1,t_2,d$ as free variables, and setting $t_1=1$, we get one of the solutions of above equations 
Again the above system admits many solutions. We pick one with $t_1 = 1, t_2 = 2, b=\frac{1}{2} $ and
\begin{align*}
    & a= \frac{1}{80} \left(3 \sqrt{21}+23\right),
    c= \frac{1}{80} \left(\sqrt{21}+41\right), \\
    & d = \frac{1}{40} \left(-7 \sqrt{\frac{3}{2}}-\sqrt{\frac{7}{2}}\right),
    y = \frac{1}{40} \left(\sqrt{\frac{3}{2}}+3 \sqrt{\frac{7}{2}}\right), \\
    & \EFB = \frac{5}{2},
    s_1 =-\frac{\sqrt{\frac{7}{2}}}{3},
    s_6 = -\frac{\sqrt{\frac{3}{2}}}{2}
\end{align*}

Therefore the CLS amplitudes and the hopping matrices are:  
\begin{align*}
    \vpsi_1 & = \left(\begin{array}{c}
        \frac{1}{40} \left(-\sqrt{\frac{3}{2}}-3 \sqrt{\frac{7}{2}}\right) \\
        \frac{1}{40} \left(\sqrt{\frac{3}{2}}+3 \sqrt{\frac{7}{2}}\right) \\
        \frac{1}{40} \left(\sqrt{\frac{3}{2}}+3 \sqrt{\frac{7}{2}}\right) \\
    \end{array}\right)\\
    \vpsi_2 & = \left(\begin{array}{c}
        \frac{1}{80} \left(3 \sqrt{21}+23\right) \\
        \frac{1}{2} \\
        \frac{1}{80} \left(\sqrt{21}+41\right) \\
    \end{array}\right)\\
    \vpsi_3 & = \left(\begin{array}{c}
        \frac{1}{40} \left(-7 \sqrt{\frac{3}{2}}-\sqrt{\frac{7}{2}}\right) \\
        \frac{1}{20} \left(-7 \sqrt{\frac{3}{2}}-\sqrt{\frac{7}{2}}\right) \\
        \frac{1}{40} \left(-7 \sqrt{\frac{3}{2}}-\sqrt{\frac{7}{2}}\right) \\
    \end{array}\right)\\
    H_1 & = \left(\begin{array}{ccc}
        -\frac{\sqrt{\frac{7}{2}}}{3} & -\frac{\sqrt{\frac{7}{2}}}{3} & 0 \\
        \frac{\sqrt{\frac{7}{2}}}{6} & \frac{\sqrt{\frac{3}{2}}}{2}+\frac{\sqrt{\frac{7}{2}}}{6} & -\frac{\sqrt{\frac{3}{2}}}{2} \\
        0 & -\sqrt{\frac{3}{2}} & \sqrt{\frac{3}{2}} \\
    \end{array}\right)\\
    H_0 & = \left(\begin{array}{ccc}
        0 & 1 & 0 \\
        1 & 0 & 2 \\
        0 & 2 & 0 \\
    \end{array}\right)
\end{align*}
which gives a flat band with energy $\EFB=5/2$. Schematics and the band structure of this lattice is shown in figure \ref{fig:u3_nu3_latt_str_example}.

\bibliography{flatband,general}

\end{document}